# Dissolution and Recrystallization Behavior of Li$_3$PS$_4$ in Different Organic Solvents

Kerstin Wissel[a,*], Luise M. Riegger[b,c], Christian Schneider[d], Aamir I. Waidha[a], Theodosios Famprikis[e], Yuji Ikeda[f], Blazej Grabowski[f], Robert E. Dinnebier[d], Bettina V. Lotsch[d,g], Jürgen Janek[b,c], Wolfgang Ensinger[a] and Oliver Clemens[h]

[a] Technical University of Darmstadt, Institute for Materials Science, Materials Analysis, Alarich-Weiss-Straße 2, 64287 Darmstadt, Germany

[b] Justus-Liebig-University Gießen, Institute for Physical Chemistry, Heinrich-Buff-Ring 17, 35392 Gießen, Germany

[c] Justus-Liebig-University Gießen, Center for Materials Research (ZfM), Heinrich-Buff-Ring 17, 35392 Gießen, Germany

[d] Max Planck Institute for Solid State Research, Heisenbergstraße 1, 70569 Stuttgart, Germany

[e] Delft University of Technology, Department of Radiation Science and Technology, Mekelweg 15, Delft 2629JB, The Netherlands

[f] University of Stuttgart, Institute for Materials Science, Materials Design, Pfaffenwaldring 55, 70569 Stuttgart, Germany

[g] Ludwig-Maximilians-Universität München, Department of Chemistry, Butenandtstraße 5-13, 81377 München, Germany

[h] University of Stuttgart, Institute for Materials Science, Chemical Materials Synthesis, Heisenbergstraße 3, 70569 Stuttgart, Germany

Corresponding Author:

Dr. Kerstin Wissel

Email: kerstin.wissel@tu-darmstadt.de

Fax: +49 6151 16-21991

**Abstract**

Solid state batteries can be built based on thiophosphate electrolytes such as β-Li$_3$PS$_4$. For the preparation of these electrolytes, various solvent-based routes have been reported. For recycling of end-of-life solid state batteries based on such thiophosphates, we consider the development of dissolution and recrystallization strategies for the recovery of the model compound β-Li$_3$PS$_4$. We show that recrystallization can only be performed in polar, slightly protic solvents such as N-methylformamide (NMF). The recrystallization is comprehensively studied, showing that it proceeds via an intermediate phase with composition Li$_3$PS$_4$·2NMF, which is structurally characterized. This phase has a high resistivity for the transport of lithium ions and must be removed in order to obtain a recrystallized product with a conductivity similar to the pristine material. Moreover, the recrystallization from solution results in an increase of the amorphous phase fraction next to crystalline β-Li$_3$PS$_4$, which results in a decrease of the activation energy to 0.2 eV compared to 0.38 eV for the pristine phase.

# 1 Introduction

Inherent safety concerns related to the use of combustible liquid electrolytes and limited energy density of commercial lithium ion batteries (LIBs) have prompted the development of different next-generation battery technologies, of which all-solid-state batteries (ASSBs) utilizing solid electrolytes (SEs) are regarded to be highly promising. Various material classes (e.g. oxides, halides and sulfides) are currently investigated for their use as SE. Among these, sulfides show some of the highest room-temperature ionic conductivities (up to ≈ $10^{-2}$ S/cm).[1] Prominent examples of these sulfide SEs are Li-P-S (LPS)-based glasses or glass ceramics (e.g. $xLi_2S\cdot(100-x)P_2S_5$, $Li_7P_3S_{11}$, β-$Li_3PS_4$), argyrodites $Li_6PS_5X$ (X = Cl, Br, I), thio-LISICONs or $Li_{11-x}M_{2-x}P_{1+x}S_{12}$ (M = Ge, Sn, Si). β-$Li_3PS_4$, $Li_7P_3S_{11}$ and argyrodites can be obtained via scalable liquid-phase syntheses. For example, β-$Li_3PS_4$ can be synthesized when its precursors $Li_2S$ and $P_2S_5$ are either dissolved or dispersed in different organic solvents. Commonly used solvents for the synthesis are e.g. acetonitrile (ACN)[2], diethylene glycol dimethyl ether (DEGDME)[3], dimethyl carbonate (DMC)[4], dimethoxyethane (DME)[5], ethyl acetate (EA)[6], ethylenediamine (EDA)[7], N-methylformamide (NMF)[8-10] or tetrahydrofuran (THF)[11]. During the synthesis, the precipitation of $Li_3PS_4$ complexed with the respective solvent (e.g. $Li_3PS_4\cdot ACN$[12], $Li_3PS_4\cdot DME$[5], $Li_3PS_4\cdot 2EA$[6], $Li_3PS_4\cdot 3THF$[11]) takes place. After an additional thermal treatment to remove the solvent, β-$Li_3PS_4$ crystallizes from these complexes.[12]

With the expected commercialization of ASSBs, sustainable battery recycling strategies will be of rising urgency. However, to date, most of the recycling research efforts focus on traditional LIBs with organic liquid electrolytes and are based on pyrometallurgical and/or hydrometallurgical processes, aiming at the recycling of the valuable metals (predominantly Ni, Co, Mn) contained in the electrode materials only. Irrecoverable losses of the liquid electrolytes and lithium lower the overall material recovery rate drastically, while the energy and cost efficiency remain low as well. As ASSBs are still under development, there is great opportunity to early on establish economically viable and efficient recycling strategies taking also sustainable battery design concepts with a focus on the recyclability of such batteries into account. Owing to the potential for solution-processing of sulfide electrolytes, sulfide-based ASSBs might offer significant potential in this respect since direct recycling concepts involving the dissolution and recrystallization of the electrolyte could be adopted.[13] This way, the insoluble electrode materials could be separated via filtration or centrifugation from the dissolved electrolyte and processed separately. So far only little is known on the recycling of sulfide electrolytes: Tan et al.[14] reported on a potential recycling strategy for thiophosphate electrolyte-based Li|$Li_6PS_5Cl$|$LiCoO_2$ cells employing ethanol as solvent allowing for a

separation of LiCoO$_2$ from the formed suspension. After the removal of the solvent Li$_6$PS$_5$Cl could be recovered.

A key requirement for this recycling strategy is that the electrolyte is fully dissolvable, while avoiding a chemical degradation of the electrolyte when in solution. In this context, it has been noted that polar solvents readily solvate thiophosphate units.[13, 15] Keeping thiophosphate units intact is essential, although it can be difficult since sulfide electrolytes show a high reactivity towards most solvents. It should be noted that dissolution strategies must be conceptionally distinguished from synthesis strategies, as not precursor materials but the sulfide electrolytes have to be processed in the solvent. For β-Li$_3$PS$_4$, there is a large number of studies on solvent-based processing in the context of the preparation of active electrode coatings, composite electrodes or separator films, however, most of these studies start from the precursor Li$_2$S and P$_2$S$_5$ (or glassy 75Li$_2$S·25P$_2$S$_5$ (mol%)) and/or use non-polar or weakly polar aprotic solvents only.[16-24] Thus, this study aims to investigate the impact of solvent treatments on β-Li$_3$PS$_4$. For this, a variety of solvents covering a broad range of physical and chemical properties is systematically screened and the solubility of β-Li$_3$PS$_4$ within them is investigated. Structural and chemical changes of the precipitated phases formed after the removal of the solvent are examined using X-ray diffraction (XRD) and Raman spectroscopy. A detailed investigation on the recrystallization behavior of β-Li$_3$PS$_4$ after complete dissolution in NMF is carried out. The formation of a previously unreported Li$_3$PS$_4$·2NMF complex is observed which is determined to have a *C*-centered monoclinic crystal structure in the space group *C*2/*c*. Effects of the dissolution in NMF on the obtained materials are investigated via a combination of X-ray and neutron diffraction, Rietveld analysis, X-ray photoelectron (XPS), infrared (IR) and electrochemical impedance (EIS) spectroscopy. Morphology changes are followed using scanning electron microscopy (SEM).

## 2 Experimental

### 2.1 Material Preparation

The pristine electrolyte β-Li$_3$PS$_4$ was purchased from NEI Corporation (USA). Material handling and preparation was carried out in inert atmosphere.

To investigate the effect of a solvent treatment on β-Li$_3$PS$_4$, different organic solvents (hexane (anhydrous, Alfa Aesar), toluene (anhydrous, 99.8 %, Sigma Aldrich), tetrahydrofuran (THF, anhydrous, 99.8+ %, Alfa Aesar), ethyl acetate (EA, 99.8 %, Sigma Aldrich), 1,2-dimethoxyethane (DME, anhydrous, 99.5 %, Sigma Aldrich), acetonitrile (ACN, anhydrous, 99.8+ %. Alfa Aesar), N-methyl formamide (NMF, 99%, thermo scientific), isopropanol (i-PrOH, anhydrous, max. water 0.003 %, VWR Chemicals), ethanol (EtOH, anhydrous, max. water 0.003 %, VWR Chemicals), and methanol (MeOH, anhydrous, 99.9%, thermo scientific) were selected. This selection covers a variety of solvents that are commonly used in literature for the synthesis of β-Li$_3$PS$_4$ (i.e., THF, EA, DME, and ACN (polar, aprotic); NMF (polar, (weakly) protic), as well as other nonpolar solvents (i.e., hexane and toluene) and polar, stronger protic solvents (i.e., i-PrOH, EtOH and MeOH). The physical and chemical properties of the used solvents are summarized in Table S1. The solvents were dried over molecular sieve (3 Å, 20% m/v, Sigma-Aldrich). The molecular sieve was removed from the solvent after 72 h via filtration. To avoid any contamination from colloidal molecular sieve particles within the solvent, vacuum distillation was carried out in addition. The water contents of the solvents were determined by Karl Fischer titration (Titrator Compact C10SX, Mettler-Toledo) and are also given in Table S1. 250 mg of the electrolyte were mixed with 50 ml solvent under Ar-atmosphere. After 4 h of stirring, the Schlenk-flasks were connected to a Schlenk line and heated to 120 °C under vacuum (p ≈ 1-3 • 10$^{-2}$ mbar) for 4 h to remove the respective excess solvent. For ACN, DME and NMF, additional heat treatments of the obtained powders were performed at 240 °C in a vacuum oven (glass oven B-585 drying, Büchi).

For further dissolution experiments with NMF, different β-Li$_3$PS$_4$ to NMF solid-to-liquid ratios were investigated. Using a ratio of 50 mg to 1 ml and a stirring time of 30 min, 3 g of β-Li$_3$PS$_4$ were dissolved in NMF and the sample was heated to a temperature of 120 °C at a pressure of ~ 2• 10$^{-2}$ mbar for 4 h. This sample was subsequently divided into several samples and each one was heated to a temperature between 140 and 240 °C under vacuum for additional 4 h.

### 2.2 Characterization

#### 2.2.1 X-ray and Neutron Powder Diffraction and Rietveld analysis

XRD patterns were recorded on a Rigaku SmartLab in Bragg-Brentano geometry with Cu K$_\alpha$ radiation with a wavelength of 1.542 Å and a Hypix-3000 detector. Samples were measured inside low background air-tight sample holders (Rigaku), which were sealed inside an Ar-filled glovebox.

Neutron powder diffraction (NPD) experiments were performed at the PEARL diffractometer of the Reactor Institute Delft (TU Delft, the Netherlands)[25]. Approximately 600 mg of sample were loaded in a 6 mm diameter can made from V-Ni null-scattering alloy which was sealed airtightly using a rubber O-ring. Handling of the powder was performed in an Ar-filled glovebox. The sample can was placed in a neutron-transparent vacuum box connected to primary vacuum (~$10^{-3}$ mbar). The diffractogram was measured over ~21 h using a wavelength of 1.667 Å selected using the (533) reflection of a Ge[511] monochromator. The instrument background determined from measurements of the empty can was subtracted from the raw diffractogram. Detector pixel normalization was performed by fitting a measurement of a PMMA rod in the same configuration.

Analysis of diffraction data was performed via the Rietveld method with the program TOPAS V.6.0. For the crystal structure determination, a coupled Rietveld analysis of X-ray and neutron powder diffraction data was performed. The instrumental intensity distribution of the XRD and NPD instruments were determined empirically from a fundamental parameter set determined using a reference scan of $LaB_6$ (NIST 660a) and $Al_2O_3$ (NIST 676a), respectively. Microstructural parameters (i.e., crystallite size and strain broadening) were refined to adjust the peak shapes.

For the determination of amorphous phase contents, the samples were mixed in a defined weight ratio with $Al_2O_3$ (calcinated at 1100 °C) and XRD patterns were recorded. The calculation of the respective amorphous fraction was performed using the internal standard method as implemented in TOPAS V.6.0.

2.2.2 Thermogravimetric and evolved gas analysis

To quantify the amount of solvent in $Li_3PS_4 \cdot x$NMF thermogravimetric analysis (TGA) was performed. Furthermore, evolved gas analysis (EGA) was performed to identify the compounds released during TG analysis. Prior to the TGA/EGA measurement, blank samples were measured to minimize buoyancy effects in the TGA and to obtain background spectra in the EGA. For the TGA measurement about 10 mg of $Li_3PS_4 \cdot x$NMF powder sample were loaded into an $Al_2O_3$ crucible and subsequently transferred into the furnace chamber. The TGA was carefully purged several times with helium to avoid atmospheric contaminates. The sample was heated with a heating rate of 10 K min$^{-1}$ to 300 °C (Netzsch Jupiter STA 449 F3, SiC furnace, type S thermocouple sample carrier, 70 ml min$^{-1}$ He (5N Air Liquid) stream). The TGA data was analyzed with the software *Proteus Analysis* (Netzsch). Sampling of the evolved exhaust gas was performed at 86 °C, 212 °C, and 278 °C. The sampled gas was transferred *via* an inert, heated line to a gas chromatograph (GC, Agilent 8890 GC System, HP-5MS UI column with 30 m · 0.25 mm · 0.25 µm, 20:1 injection split ratio, sample inlet temperature 300 °C, He carrier with 1.2 mL min$^{-1}$). The GC oven was set to a constant temperature of 100 °C. After passing the GC, the separated components were detected in a mass spectrometer (MS, Agilent 5977B GC/MSD, EI mode, 230 °C ion source temperature, 150 °C quadrupole temperature). The total ion current of the GC peaks was measured. The mass spectra were analyzed by means of best matching database entries (NIST database, *MSD ChemStation Data Analysis*).

### 2.2.3 DFT calculations

To further support the atomic positions determined in the Rietveld analysis, structural optimization based on ab initio density functional theory (DFT) was also conducted. The optimization was done for the primitive cell including 52 atoms, where the primitive lattice vectors are (**a** − **b**)/2, (**a** + **b**)/2, and c. The DFT calculations were performed using the VASP code. [26-28] with the plane-wave basis projector augmented wave (PAW) method. [29] The exchange–correlation energy was obtained within the generalized gradient approximation (GGA) of the Perdew–Burke–Ernzerhof (PBE) form. [30] The plane-wave cutoff energy was set to 520 eV. Reciprocal spaces were sampled by a Γ-centered 6 × 6 × 3 k-point mesh and the tetrahedron method with the Blöchl correction. [31] H 1s, Li 2s, C 2s2p, N 2s2p, O 2s2p, P 3s3p, and S 3s3p orbitals were treated as the valence states. Total energies were minimized until they converged within 1·10$^{-6}$ eV per simulation cell for each ionic step. Cell volume, cell shape, and internal atomic positions were optimized so that the forces on atoms and the stress components on the unit cell became less than 1·10$^{-2}$ eV/Å and 1·10$^{-4}$ eV/Å$^3$, respectively.

### 2.2.4 Scanning electron micrcopy

SEM images were recorded using a secondary electron detector of a FEI Quanta 250 SEM operating at 30 keV. Prior to the measurements, a layer of Au was sputtered onto the samples.

### 2.2.5 Raman spectroscopy

Raman spectra were recorded with a confocal micro-Raman spectrometer Horiba HR 800 equipped with a laser wavelength of 532 nm. For calibration of the spectrometer, the 521 cm$^{-1}$ Stokes signal of a silicon wafer was used. Calibration measurements were performed after each measurement. Samples were measured in glass capillaries sealed under inert atmosphere.

### 2.2.6 Infrared spectroscopy

Fourier-transform infrared spectroscopy (FTIR) measurements were conducted on a Varian spectrometer. Samples were characterized via attenuated total reflection (ATR) using an ATR unit (Specac) equipped with a reactive sample anvil. For the sample preparation, the ATR unit was transferred into an Ar-filled glovebox, where the sample was compressed under inert atmosphere, sealing the powder from the environment with an O-ring. To limit the duration of possible exposure to air, the measurement was conducted as quick as possible after the unit was transferred out of the glovebox.

### 2.2.7 X-ray photoelectron spectroscopy

XPS measurements were carried out on a PHI 5000 VersaProbe II Scanning ESCA Microprobe (Physical Electronics GmbH) with a monochromatized Al Kα source (1486.6 eV). The beam had a power of 50 W and a diameter of 200 µm. The sample surface was charge neutralized with slow argon-ions and electrons. $Ar^+$ ions accelerated with 0.5 kV were used to sputter a depth profile. For the detailed spectra an analyzer pass energy of 46.95 eV, a step time of 50 ms and a step size of 0.2 eV were used. CasaXPS software (Casa Software Ltd) was used for data analysis with a Shirley-type background correction and a GL(30) line shape. Before fitting, all the peaks were calibrated to the binding energy of adventitious $sp^3$-carbon (284.8 eV) first. To avoid detrimental surface effects the spectra were then calibrated to the binding energy of the S2p main component $PS_4^{3-}$ at 161.7 eV. [32, 33]

### 2.2.8 Electrochemical impedance spectroscopy

The conductivity of the obtained materials was measured under pressure in a CompreDrive (rhd instruments). For this, 80 to 100 mg of each powder was loaded into a measuring cell. Carbon-coated aluminum electrodes were used to ensure contact. After applying 380 MPa for 3 min, a constant pressure of 50 MPa was applied. Electrical impedance measurements were performed using an electrochemical impedance analyzer NEISYS (Novocontrol Technologies) in a frequency range between 7 MHz and 1 Hz with an amplitude of 10 mV in a temperature range between -30 and 100 °C. Received data was analyzed using the software RelaxIS3 (rhd instruments GmbH & Co. KG).

## 3 Results and Discussions

### 3.1 Solvent treatment of β-Li$_3$PS$_4$ in different organic solvents

To study the impact of a solvent treatment on β-Li$_3$PS$_4$, a selection of different organic solvents has been made based on their different physical and chemical properties (Table S1), covering a spectrum from non-polar over polar aprotic to polar protic solvents. As can be seen in Figure 1 a), only the polar protic solvents NMF, i-PrOH, EtOH and MeOH are able to form a clear solution containing a significant quantity of β-Li$_3$PS$_4$ (fixed ratios of 5 mg β-Li$_3$PS$_4$ per 1 ml of solvent were investigated). However, it should be noted that, while β-Li$_3$PS$_4$ is dissolved in less than 2 minutes in EtOH and MeOH, incomplete dissolution was observed in the case of i-PrOH and the mixture remains slightly cloudy.

For all other non-polar and polar aprotic solvents, no to negligible solubility was found leading to the presence of considerable amounts of white precipitates. A sedimentation of the particles takes place within 1 to 2 minutes showing that stable suspensions cannot be formed. The blue color of the ACN suspension points to a certain reactivity between β-Li$_3$PS$_4$ and ACN under the formation S$_3^-$ radicals [34, 35], while colorless supernatants were observed for the other non-polar and polar aprotic solvents indicating no to minor reactions between sulfur-containing species and the solvents [36]. The solutions with polar protic solvents have a pale-yellow color.

After the evaporation of the solvents at 120 °C and reduced pressures, the obtained precipitates were collected and X-ray diffraction and Raman spectroscopy measurements (Figure 1 b and c) were performed. The orthorhombic crystal structure of β-Li$_3$PS$_4$ is recovered for all solvents except for i-PrOH, EtOH and MeOH. It should be noted that the pristine β-Li$_3$PS$_4$ contains ~ 2 wt.% Li$_2$S as an impurity phase, which is also present in the recrystallized samples. The phase fraction of Li$_2$S remains constant in all samples.

As has been reported previously when synthesizing β-Li$_3$PS$_4$ from DME and ACN[5, 12], insoluble Li$_3$PS$_4$·ACN and Li$_3$PS$_4$·DME complexes are obtained when using a moderate drying temperature of 120 °C (Figure S1). Only after heating to higher temperatures, these complexes decompose and the formation of β-Li$_3$PS$_4$ takes place. A similar behavior is found for NMF suggesting the formation of a so far unknown NMF complex which also transforms to β-Li$_3$PS$_4$ at elevated temperatures. The detailed crystal structure and property characterization of this phase is described in section 3.2. For the alcohols, on the other hand, partial decomposition under the formation of Li$_2$S and other unknown phases and/or amorphization is observed which is also in agreement with previous studies. [15, 37, 38] Interestingly, alcohols can be used to dissolve and recrystallize other sulfide electrolytes such as argyrodites Li$_6$PS$_5$X [38-40] or Li$_7$PS$_6$ [37], even though they also contain PS$_4^{3-}$ units. For these electrolytes, the presence of excess Li$_2$S and/or LiX seems to play an important role for the stabilization of PS$_4^{3-}$ units during the dissolution process. This shows, however, that it is not necessarily possible to dissolve different electrolytes (simultaneously) in a specific solvent which could become important with respect to potential large-scale recycling considering different thiophosphate-based cell chemistries. Therefore, an additional sorting and separation step would be required. Furthermore, this illustrates that for each sulfide electrolyte which could find application in future ASSB cells the dissolution behavior has to be studied individually.

The Raman spectra are in excellent agreement with the diffraction data. The signals at ~ 421 cm$^{-1}$ dominating the Raman spectra correspond to the symmetric stretching vibration of ortho-thiophosphate units $PS_4^{3-}$ [41, 42] which are expected to be present in the samples showing the β-$Li_3PS_4$ structure. The decrease of this signal and the appearance of other signals (e.g. polysulfide species in the range ~ 450 to 500 cm$^{-1}$, $P_2S_6^{4-}$ at ~ 390 cm$^{-1}$)[38] for the alcohols demonstrates that these samples decomposed significantly. This also confirms findings from previous studies. [36-38]

It can be concluded that only NMF dissolves β-$Li_3PS_4$ under the retention of the $PS_4^{3-}$ units, making a recrystallization of the crystal structure possible. Several factors seem to play an important role: NMF is a strongly polar, weakly protic solvent containing NH bonds. The polarity and the weakly positively polarized hydrogen seem to be necessary to solvate the $PS_4^{3-}$ anionic species. In comparison, the alcohols possess OH bonds which make them more nucleophilic. Though this leads also to the solvation of the anion, it can also cause decomposition under the formation of $PS_{4-x}(OR)_x$ at the same time since PO bonds are thermodynamically more stable.[15] This reaction step is irreversible and prevents the recrystallisation of β-$Li_3PS_4$ or $Li_3PS_4$·solvent complexes. The fact that i-PrOH does not dissolve β-$Li_3PS_4$ completely might also show that not only the OH or NH bonds, respectively, play an important role in the dissolution process but also the size of the solvent molecules. Due to the size of i-PrOH the nucleophilic substitution reaction can be expected to be kinetically hindered [15]. Additionally, the polarity could play a role since NMF is much more polar than i-PrOH.

Non-polar and aprotic solvents, on the other hand, cannot break the bonding between the Li$^+$ cations and $PS_4^{3-}$ anions. Consequently, this does not lead to a complete dissolution but to the formation of suspensions. From these, β-$Li_3PS_4$ can be recovered after the removal of the solvent, however, they do not have a significantly high solubility for solvating a considerable amount of β-$Li_3PS_4$.

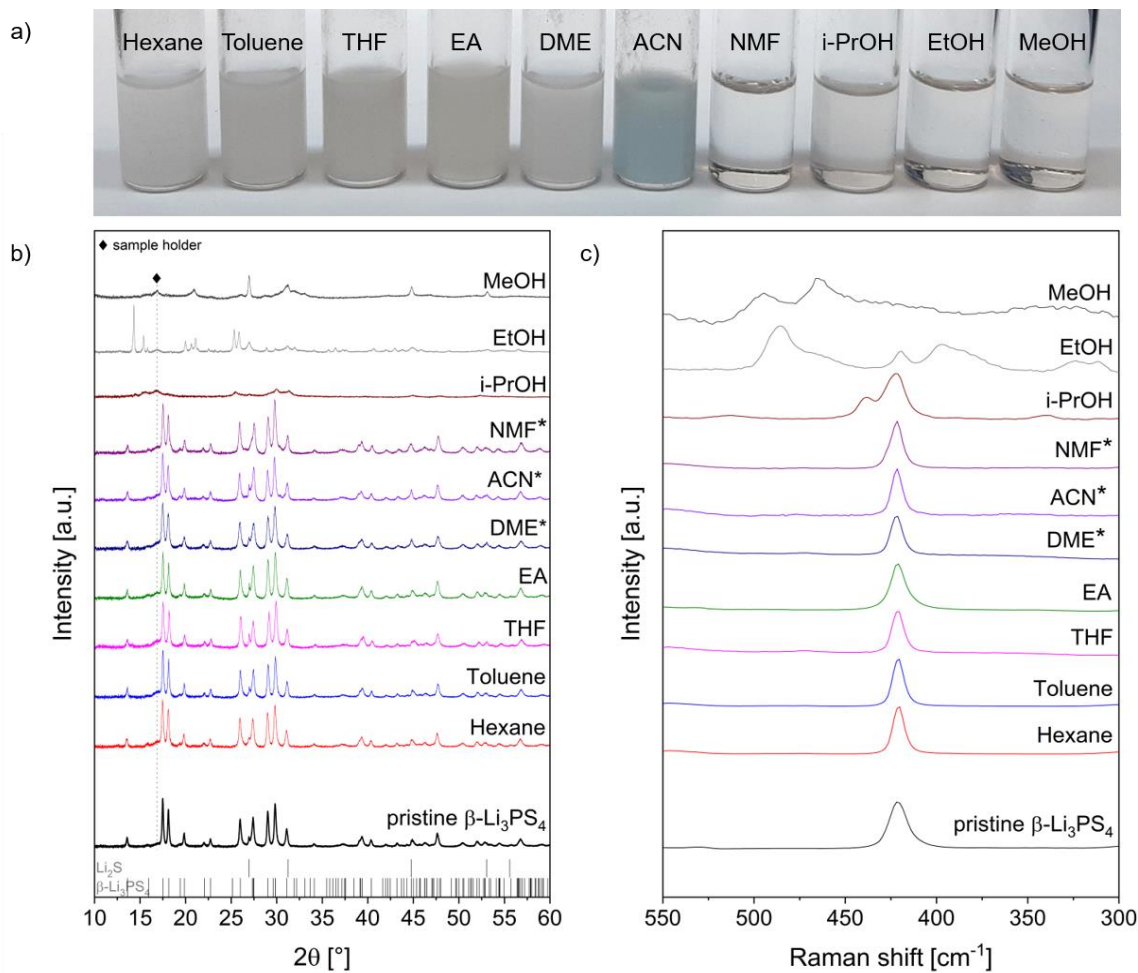

*Figure 1: a) Photograph of β-Li₃PS₄ in different organic solvents; b) and c) X-ray diffraction patterns and Raman spectra of dried precipitates obtained after solvent treatment of β-Li₃PS₄ in organic solvents in comparison to pristine β-Li₃PS₄. To remove the solvents, the samples were heated to 120 °C for 4 h at reduced pressures. To decompose Li₃PS₄·solvent complexes (Figure S1), samples marked with * were additionally heated to 240 °C for 4 h at reduced pressures.*

### 3.2 Dissolution of β-Li₃PS₄ in NMF and its recrystallization mechanism

A complete dissolution of β-Li$_3$PS$_4$ in NMF is already observed when using a molar ratio of ~ 1:14 of β-Li$_3$PS$_4$ to NMF which corresponds to an approximate solid to liquid ratio of 200 mg per 1 ml. For this highest concentration, the obtained clear solution (Figure S2) has a bright yellow color and is highly viscous. Lower relative amounts of NMF cannot fully dissolve β-Li$_3$PS$_4$ and a cloudy suspension is formed. An increase of the amount of NMF by a factor of ~ 4 leads, on the other hand, to a significant decrease of the viscosity of the obtained solution. With this, the time needed for the dissolution also drops considerably from ~ 2 h to 2 to 3 min. This lower viscosity is important when considering that in a recycling process a separation of the electrode materials from the electrolyte solution should by performed using filtration or centrifugation.

The boiling point of NMF has been reported to be between 180 and 200 °C. [43, 44] As described in section 3.1, when using a lower evaporation temperature of 120 °C, the removal of excess NMF from the solution results in the formation of a white powder with an X-ray powder diffraction pattern significantly different to β-$Li_3PS_4$ (Figure 2 a) for which the crystal structure has not been reported previously. This phase can, however, be transformed into β-$Li_3PS_4$ after additional heat treatments. For this, the sample was heated to various temperatures between 140 and 240 °C (Figure 2 a). A quantitative phase analysis (Figure 2 b) reveals that temperatures significantly above the boiling point of NMF are required to recrystallize a considerable fraction of β-$Li_3PS_4$, suggesting that a certain driving force is required to transform the intermediate phase to β-$Li_3PS_4$. Therefore, it can be assumed that this intermediate phase is a $Li_3PS_4 \cdot xNMF$ complex phase as also observed when using other solvents like for example ACN[12] or DME[5]. Only after heating to 240 °C, phase-pure samples are obtained, whereas for all other temperatures phase mixtures of the unknown phase and β-$Li_3PS_4$ are found. When comparing the XRD patterns of the pristine and the recrystallized β-$Li_3PS_4$, no significant differences in reflex intensity ratios are found. Rietveld analysis indicates a small change of lattice parameters for the pristine and recrystallized β-$Li_3PS_4$ ($a_{pristine}$ = 12.9959(4) Å, $b_{pristine}$ = 8.0504(3) Å, $c_{pristine}$ = 6.1430(2) Å vs. $a_{recrystallized}$ = 12.9564(5) Å, $b_{recrystallized}$ = 8.0961(3) Å, $c_{recrystallized}$ = 6.1428(2) Å). The cell volume of the recrystallized β-$Li_3PS_4$ is bigger by only 0.26 %, suggesting only minimal structural changes (e.g. difference in defect concentrations). Additionally, an increase of the amorphous phase fraction from ~ 8 wt.% in the pristine sample to ~ 35 wt.% in the recrystallized sample was determined using the internal standard method (Figure S3), indicating that the heating temperature of 240 °C is not high enough and/or the duration of 4 h not long enough to obtain a complete recrystallization.

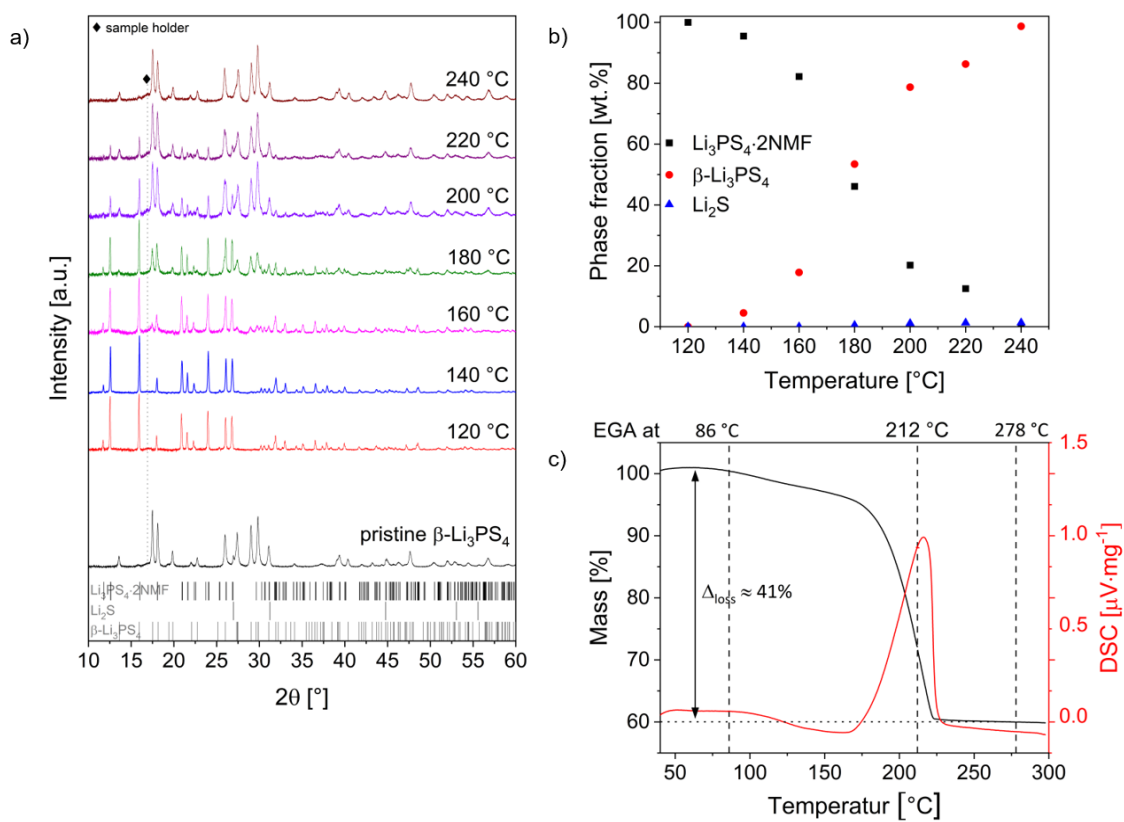

*Figure 2: X-ray diffraction patterns (a) and quantitative phase analysis (b) of re-heated samples in comparison to pristine β-Li$_3$PS$_4$ and Li$_3$PS$_4$·2NMF obtained after solvent removal at 120 °C after the treatment of β-Li$_3$PS$_4$ in NMF. The samples were heated to temperatures between 140 and 200 °C for 4 h at reduced pressures. c) TG analysis of Li$_3$PS$_4$·2NMF. EG analysis was performed at 86, 212 and 278 °C to investigate which gases are released during the TG analysis. These results are given in Figure S4.*

To understand the changes observed upon recrystallization of β-Li$_3$PS$_4$ from the intermediate phase and the overall recrystallization mechanism better, the intermediate phase was investigated in more detail and its crystal structure was determined. Indexing of the XRD pattern (Figure 3 a) indicated the formation of a *C*-centered monoclinic phase with lattice parameters of $a \approx 15.89$ Å, $b \approx 6.02$ Å, $c \approx 16.95$ Å and $β \approx 117.75$ °. A weight loss of ~ 41 % in TG analysis (Figure 2 c) shows further that this phase contains approximately 2 NMF molecules per Li$_3$PS$_4$ unit suggesting the formation of a complex with composition Li$_3$PS$_4$·2NMF. Evolving gas analysis (Figure S4) confirms that the mass loss is due to the release of NMF from the crystal structure. Moreover, taking space requirements of Li$_3$PS$_4$ [45, 46] and NMF [44, 47] into consideration, it can be concluded that the cell is likely to contain 4 formula units per unit cell.

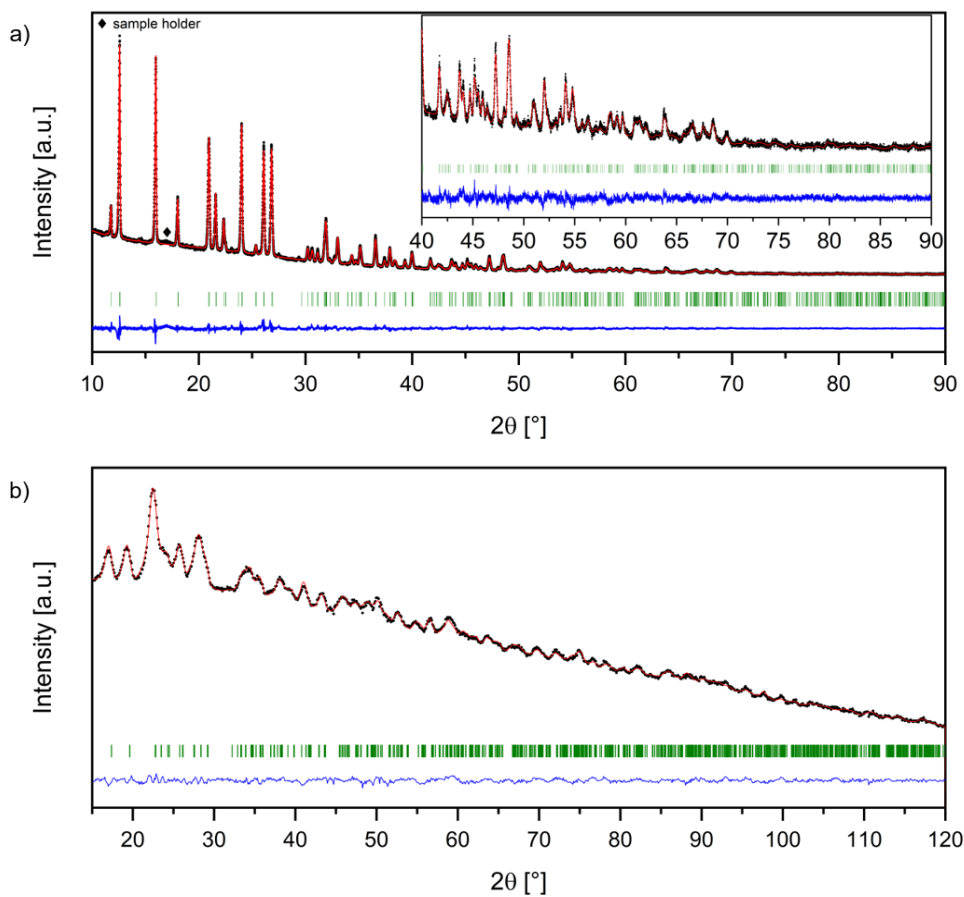

*Figure 3: Rietveld refinements of X-ray (a) and neutron (b) diffraction data of Li₃PS₄·2NMF with the model obtained using the rigid body method and simulated annealing. The inset in a) shows a magnification of the angular range between 40 and 90°. The comparably low signal-to-noise ratio of the neutron diffraction data is due to incoherent scattering of the hydrogen atoms.*

Ab initio structural solution was attempted using the rigid body method in combination with simulated annealing within different *C*-centered monoclinic structures (e.g. *C2/m* and *C2/c*). For this, rigid bodies of PS$_4$ units (as also indicated from Raman spectroscopy) as well as molecular NMF units were defined based on reported bond lengths and bond angles. [44-47] Reasonable flexibility in these parameters was given. Translation, rotation and torsion angles were refined to identify the relative orientations of the different building units. Structural solutions in *C2/m* do not result in a reasonable relative orientation of PS$_4$ units with sufficient quality of the fits. In contrast, structural solutions in *C2/c* resulted in a plausible orientation of PS$_4$ and NMF units towards each other, in agreement with the observed systematic extinctions.

By coupling additional neutron diffraction data (Figure 3b), plausible approximate positions of the Li ions could be identified, which are located on two different crystallographic sites (4*e* and 8*f*). The corresponding structural parameters after refining the positional and thermal parameters within a Rietveld analysis are given in Table 1. The obtained high thermal parameters of the H ions of the methyl group (H3-H5) indicate a partial disorder in that region of the structure, most likely explained by the rotational degree of freedom of the methyl group. The crystal structure is represented in Figure 4.

*Table 1: Structural parameters of Li$_3$PS$_4$·2NMF (space group: C2/c) as obtained from Rietveld analysis using the rigid body method and simulated annealing.*

| Atom | Wyckoff position | x | y | z | Occupancy | B [Å$^2$] |
|---|---|---|---|---|---|---|
| **Li1** | *4e* | **0** | 0.339(3) | **3/4** | **1** | 4.0(5) |
| **Li2** | *8f* | 0.820(1) | 0.196(3) | 0.059(1) | **1** | 9.4(6) |
| **P1** | *4e* | **0** | 0.8516(6) | **3/4** | **1** | 3.1(1) |
| **S1** | *8f* | 0.8801(2) | 0.0472(3) | 0.7083(2) | **1** | 3.1(1) |
| **S2** | *8f* | 0.0137(2) | 0.6457(3) | 0.8528(1) | **1** | 3.8(1) |
| **C1** | *8f* | 0.6128(5) | 0.747(1) | 0.0228(4) | **1** | 4.8(1) |
| **C2** | *8f* | 0.691(4) | 0.422(5) | 0.105(2) | **1** | 4.8(1) |
| **H1** | *8f* | 0.704(8) | 0.328(8) | 0.160(3) | **1** | 4.7(3) |
| **H2** | *8f* | 0.619(2) | 0.620(3) | 0.136(3) | **1** | 4.7(3) |
| **H3** | *8f* | 0.663(5) | 0.72(1) | 0.006(5) | **1** | 20.4(8) |
| **H4** | *8f* | 0.619(7) | 0.899(6) | 0.045(2) | **1** | 20.4(8) |
| **H5** | *8f* | 0.556(4) | 0.734(9) | 0.965(3) | **1** | 20.4(8) |
| **N1** | *8f* | 0.634(2) | 0.598(3) | 0.095(1) | **1** | 4.8(1) |
| **O1** | *8f* | 0.722(4) | 0.368(7) | 0.055(2) | **1** | 4.8(1) |
| | | | | | | |
| *a* [Å]  15.8943(4) | *b* [Å]  6.0274(2) | | *c* [Å]  16.9540(4) | | β [°]  117.747(2) | |
| R$_{wp}$ (XRD+NPD) [%]  1.26 | GOF(XRD+NPD)  1.09 | | | | R$_{Bragg}$ [%]  1.69 (XRD)  0.25 (NPD) | |

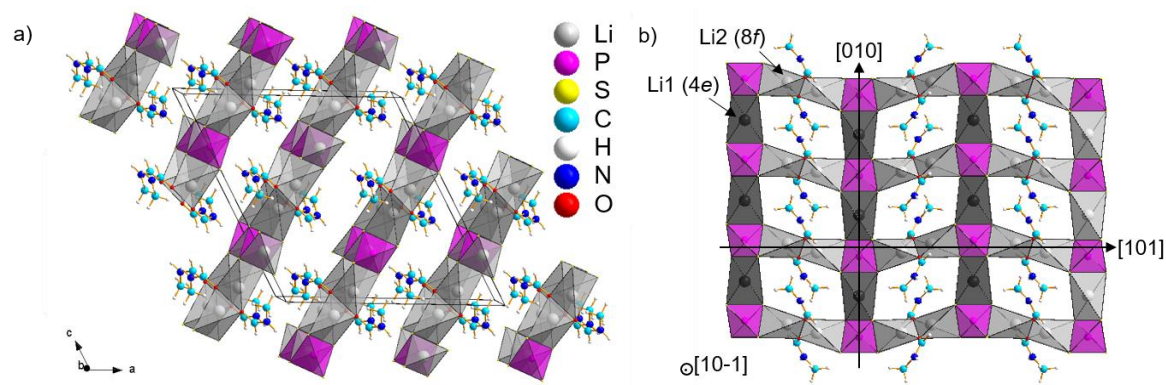

*Figure 4: Crystal structure of Li$_3$PS$_4$·2NMF (a) and top view onto one of the Li$_3$PS$_4$·2NMF layers (b) as obtained from Rietveld analysis using the rigid body method and simulated annealing.*

Additional ab initio DFT simulations confirm the validity of the determined structure. The DFT-optimized structural parameters are given Table S2 for comparison. Most interestingly, they confirm the orientation of the methyl group as well as positioning of the Li2 atom towards the oxygen group of the NMF molecule.

The structure consists of shifted layers of composition Li$_3$PS$_4$·2NMF stacked along the [10-1] direction. Two different interactions of the molecular species and Li$_3$PS$_4$ can be identified: The oxygen ions are coordinated to the Li ions on the 8*f* site and the NH groups have a weak interaction with the nearest PS$_4$ units in the adjacent layers. This results in strong ionic interactions within the layers and weaker dipole interactions perpendicular to them. The latter interaction is also reflected in a change of the NH stretching vibration IR mode (Figure 5 c). This is also important considering the solution mechanism and confirms the role of the NH bonds for stabilizing the PS$_4^{3-}$ units in solution.

Since the Li ions on the 8*f* site are connected to both PS$_4$ tetrahedra and NMF molecules, this results in chains according to …S$_2$LiO$_2$LiS$_2$PS$_2$LiO$_2$LiS$_2$… along the [101] direction. The Li ion on the 4*e* site is located in a tetrahedral LiS$_4$ polyhedron between two PS$_4$ tetrahedra. This results in chains according to …S$_2$LiS$_2$PS$_2$LiS$_2$… along the b direction. This interconnectivity of tetrahedra is also highlighted in Figure 4 b).

It becomes also evident that potential interstitial sites within and between Li$_3$PS$_4$·2NMF layers are in close distance to the non-polar methyl groups of the NMF molecules. Therefore, their occupation by ionic species such as Li$^+$ becomes less likely due to mainly non-ionic weak van-der-Waals forces around them.

When comparing the structure of Li$_3$PS$_4$·2NMF to Li$_3$PS$_4$·DME [5] and Li$_3$PS$_4$·ACN [12], significant differences can be found. The latter two compounds crystallize in the tetragonal crystal system and consist of alternating layers of Li$_2$PS$_4^-$ separated by layers with composition of (solv)$_n$Li$^+$. In contrast to Li$_3$PS$_4$·2NMF, these two structures are more similar to the structure of β-Li$_3$PS$_4$ also consisting of Li$_2$PS$_4^-$ layers separated by Li$^+$ layers. As the structural transition to the complex can be considered as a topotactic solvent insertion under volume increase, a complete dissolution of β-Li$_3$PS$_4$ in DME or ACN is not possible.

Morphological and structural changes of these samples were additionally investigated using a combination of scanning electron microscopy and Raman, infrared as well as X-ray photoelectron spectroscopy (Figure 5 and Figure 6). The morphologies of pristine β-Li$_3$PS$_4$ and Li$_3$PS$_4$·2NMF are similar with respect to the particle size, however, the particle shapes differ. Li$_3$PS$_4$·2NMF particles appear to be less regularly shaped and flakier compared to β-Li$_3$PS$_4$. This agrees with the weak dipolar interactions between the layers of Li$_3$PS$_4$·2NMF (Figure 4), leading to a comparatively easy peeling off of layers. Recrystallized β-Li$_3$PS$_4$ consists of particles with a wider particle size distribution with overall smaller particles. This is probably related to the outgassing of NMF in a random fashion from the crystal structure, resulting in stress fractions of the particles and the formation of grain boundaries.

A comparison between the Raman and FTIR spectra of pristine β-Li$_3$PS$_4$, Li$_3$PS$_4$·2NMF and recrystallized pristine β-Li$_3$PS$_4$ gives further insights into structural changes. Raman spectroscopy (Figure 5 b) confirms that all samples contain ortho-thiophosphate units indicated by the signal at ~ 421 cm$^{-1}$ [41, 42]. Other thiophosphate species cannot be observed, neither during the dissolution process (Figure S5 a), nor in Li$_3$PS$_4$·2NMF or the recrystallized β-Li$_3$PS$_4$ after the removal of NMF. The PS$_4^{3-}$ units are also preserved in solution. For Li$_3$PS$_4$·2NMF, less pronounced signals are additionally found at ~ 315 and 388 cm$^{-1}$ which can be assigned to the NMF within the structure. [9] The most significant differences in the FTIR spectra (Figure 5 c) can be again ascribed to the presence of NMF in the samples. A comparison of the spectra of pure NMF, β-Li$_3$PS$_4$ dissolved in NMF and Li$_3$PS$_4$·2NMF (Figure S5 a) reveals characteristic shifts of NMF vibration modes which dominate the spectra as the pristine β-Li$_3$PS$_4$ features no strong signals. As already mentioned, the strongest shift is observed for the NH stretching vibration mode in the range between 3100 and 3480 cm$^{-1}$ which can be attributed to the interactions between NH groups of NMF molecules and PS$_4^{3-}$ units in the layered structure of Li$_3$PS$_4$·2NMF.[48] Other modes (e.g. the CH$_3$ symmetric bend at ~ 1375 cm$^{-1}$) show less pronounced shifts since weaker interactions take place. Similar trends in band shifting are observed in the Raman spectra. The NMF related signals in the FTIR spectra of the recrystallized β-Li$_3$PS$_4$ decrease significantly, however, there seems to be still a smaller fraction of NMF present in the sample. It is unlikely that these traces remain in the crystal structure of the recrystallized sample as its presence is not indicated with respect to the cell volume of the crystalline phase formed. Additionally, the TG and EG analysis shows that NMF is released at temperatures < 240 °C used for drying of the recrystallized sample. NMF might, however, be present as surface adsorbates or in amorphous phases. In addition, the spectra of pristine and recrystallized β-Li$_3$PS$_4$ feature two signals at ~ 1042 and 961 cm$^{-1}$, which could indicate the formation of small amounts of PS$_{4-x}$O$_x^{3-}$ groups.[49] However, there is no indication for such PS$_{4-x}$O$_x^{3-}$ species in Raman spectroscopy [50, 51] and XRD (at least not in the form of crystalline phases). This also illustrates that special attention has to be paid to the amorphous phase fraction present in the sample since subtle changes in the composition of the amorphous phase (i.e., formation of phases other than amorphous Li$_3$PS$_4$) could have a significant influence on the properties of the recrystallized phases.

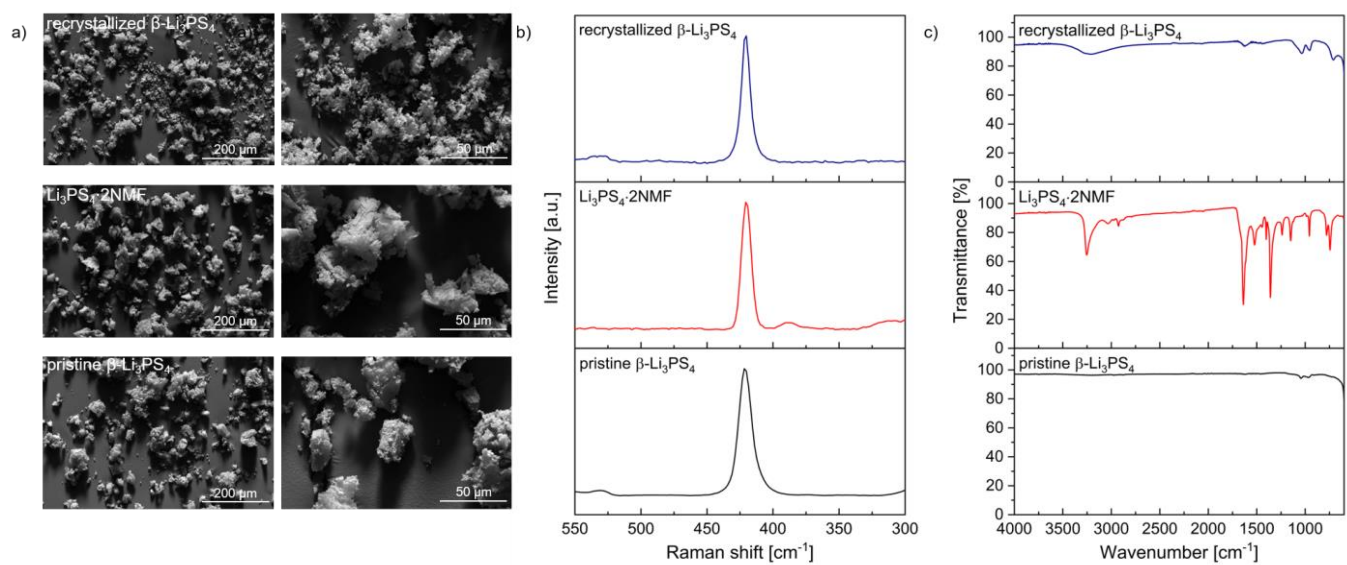

*Figure 5: Comparison of scanning electron micrographs (a), Raman (b) and FTIR (c) spectra of prisitne β-Li$_3$PS$_4$, Li$_3$PS$_4$·2NMF and recrystallized β-Li$_3$PS$_4$.*

Since the XRD analysis and Raman and FTIR spectroscopy do not show considerable structural changes between pristine and recrystallized β-Li$_3$PS$_4$, XPS measurements were additionally carried out. The comparison between the P2$p$ and S2$p$ spectra of the pristine and recrystallized β-Li$_3$PS$_4$ are shown in Figure 6. The main signal corresponds to the P-S bond in the PS$_4^{3-}$ units of β-Li$_3$PS$_4$ at 132.0 eV (P2$p$) and 161.7 eV (S2$p$).[52-54] The low-intensity signals at lower binding energies than the PS$_4^{3-}$ peaks can be assigned to reduced phosphorous species Li$_x$P (0 <x <3) and Li$_2$S in the P2$p$ and S2$p$ spectra, respectively. [53, 55] The latter agrees with the Li$_2$S impurity phase observed in the XRD measurements. The presence of P–[S]$_n$–P type bonds is indicated from the doublet at ∼ 163.0 and 164.2 eV in the S2$p$ spectrum. [55] An unambiguous assignment of the signal found at higher binding energies in the P2$p$ spectrum is not possible since P–[S]$_n$–P as well as oxygenated phosphorous species (phosphates, metaphosphates or PS$_{4-x}$O$_x^{3-}$)[32, 56] can be present at these binding energies impeding the deconvolution. For the recrystallized β-Li$_3$PS$_4$, a small shift to higher energies as well as a small increase in concentration of this signal is found. This indicates that oxygenated phosphorous species, possessing higher binding energies than P–[S]$_n$–P [57, 58], might have formed to a minor extent in addition to P-[S]$_n$–P. Most likely, these species can be assigned to the amorphous phase fraction since additional reflections are absent in the diffraction data. Besides this, an increase in the intensity of carbonate-related signals in the C 1$s$ and O 1$s$ spectra is observed in the recrystallized sample compared to the pristine sample. Nevertheless, no significant changes are observed suggesting minor degradation of the recrystallized β-Li$_3$PS$_4$ in comparison to the pristine β-Li$_3$PS$_4$ after the solvent treatment.

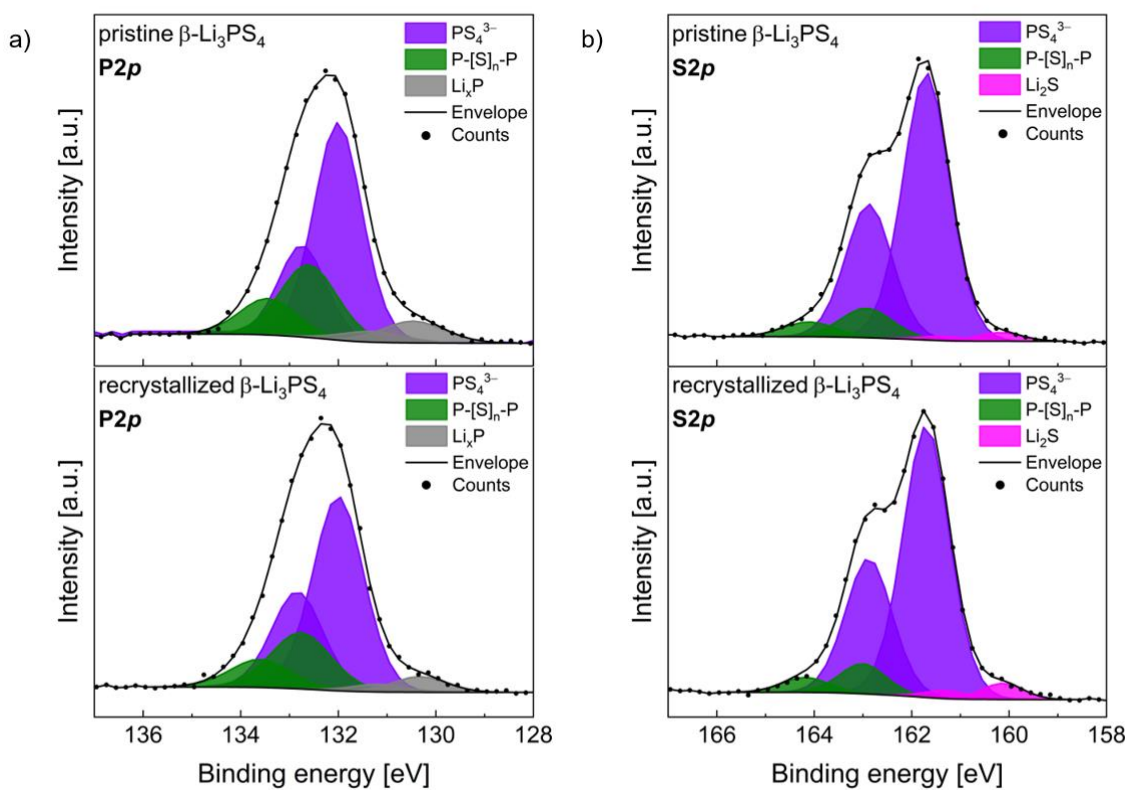

*Figure 6: Comparison between P2p (a) and S2p (b) XPS spectra of pristine and recrystallized β-Li₃PS₄.*

To investigate the influence of the solvent treatment on the ionic conductivity, impedance measurements were performed on pristine β-Li$_3$PS$_4$, Li$_3$PS$_4$·2NMF and recrystallized β-Li$_3$PS$_4$. The Nyquist plot of the pristine and recrystallized β-Li$_3$PS$_4$ (measured at 25 °C) is shown in Figure 7 a). Both samples show at first glance a similar behavior with a half semi-circle and an x-axis intercept at high frequencies and a mid- and low-frequency tail. The Bode plots (Figure S6) indicate that a single Li-conducting phenomenon is dominant. It was found that a single, parallel R/CPE element (R=resistor, CPE=constant phase element) connected in series to a CPE can be used for fitting the Li-ion transport process and electrode-ion-blocking effect at the electrode, respectively. From the fits, the conductivity values of pristine and recrystallized β-Li$_3$PS$_4$ were determined to be $2.34 \cdot 10^{-5}$ S·cm$^{-1}$ and $2.18 \cdot 10^{-5}$ S·cm$^{-1}$ at 25 °C, respectively, which is the same within experimental error.[59] The capacitance values for the pristine and recrystallized β-Li$_3$PS$_4$ are with $3 \cdot 10^{-11}$ F vs. $8 \cdot 10^{-11}$ F, respectively, slightly different, though in the same order of magnitude. Remarkably, the activation energies of both samples are with 0.38 and 0.20 eV significantly different (Figure 7 c). The lower activation energy of recrystallized β-Li$_3$PS$_4$ could be assigned to a higher amount of amorphous phase fraction in the recrystallized phase, which was reported to be beneficial for the bulk transport of Li$^+$ ions.[60] On the other hand, the higher capacitance observed for the recrystallized sample also indicates an increased influence of grain boundaries, which might be related to slightly different surface properties (e.g., induced by solvent residuals) and chemical species being present after the recrystallization process. The asymmetry observed in the Bode plot of the recrystallized sample (Figure S6) likely indicates a more complex conduction behavior beyond a single transport process. However, due to similar time constants, these processes cannot be well deconvoluted from each other.

On the contrary, $Li_3PS_4·2NMF$ shows a completely different response. From the Nyquist plot (Figure 7 b), only a single semicircle could be observed within the high and mid frequency range along with a low frequency intercept on the x-axis and no blocking tail, which indicates that this phase is considerably less Li-ion conducting with an insufficient conductivity for an application within a solid-state battery. This can further be confirmed from the Bode plots (Figure S6), where at low frequencies the phase angle is found to be closer 0 °, indicating a resistive instead of a capacitive behavior. To obtain a good estimate of the overall resistance, two R/CPE elements were used in series. From this, a conductivity value of $1.66·10^{-9}$ S·cm$^{-1}$ at 25 °C was calculated, which is 4 orders of magnitude lower than that of the pristine and the recrystallized β-$Li_3PS_4$. This decreased ion conductivity is expected with respect to the crystal structure of $Li_3PS_4·2NMF$ since there are no continuous pathways for Li-ion conduction, and due to strong ionic interactions between lithium ions and the oxygen ions of the solvent molecules. The significantly higher activation barrier of 0.70 eV (Figure 7 c) is again in agreement with the absence of additional interstitial lithium sites within the compound. It can be deduced that any remaining crystalline $Li_3PS_4·2NMF$ would have a detrimental impact on the conductivity of recrystallized samples after a solvent treatment in a possible recycling application. Therefore, it is essential to optimize the solvent removal procedure in order to avoid highly resistive residuals which would lead to a reduced electrolyte performance. At the same time, it might be beneficial to find a compromise between the formation of recrystallized and amorphous phases fractions in order to obtain optimized conductivities.

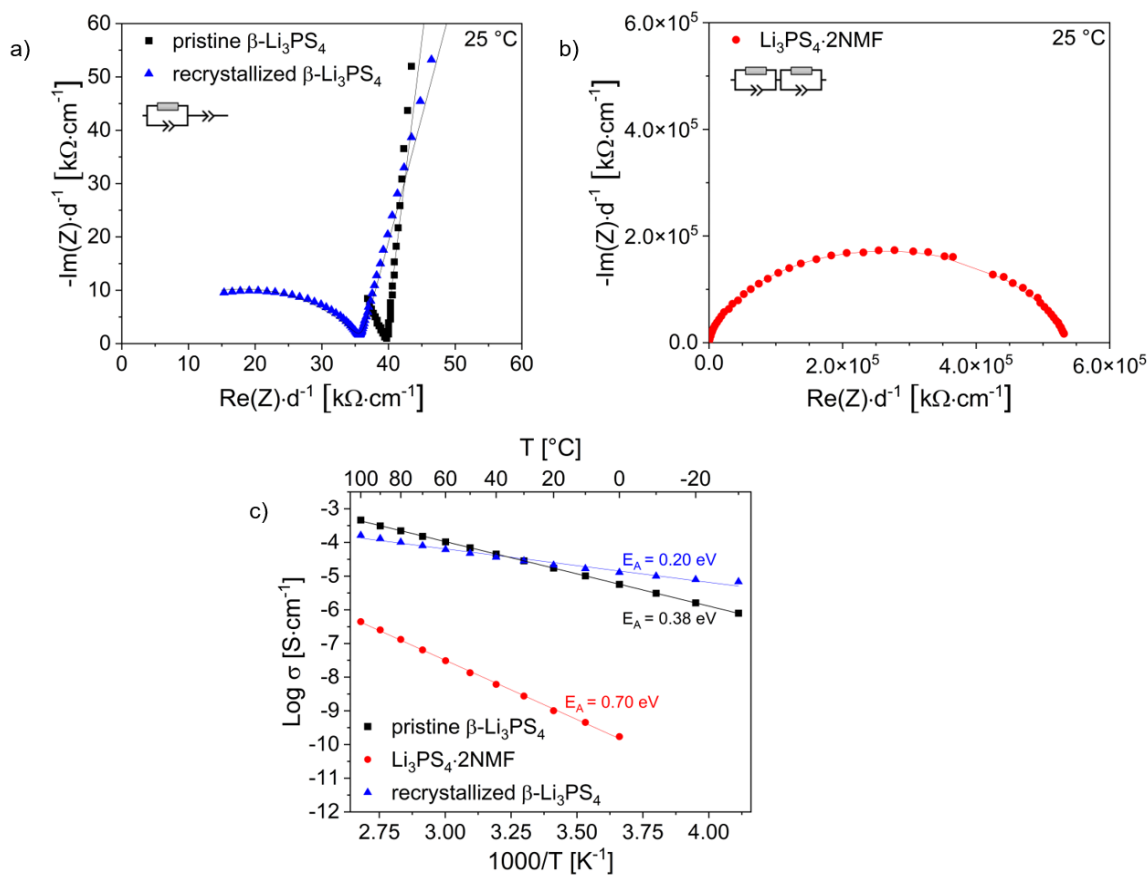

*Figure 7: Nyquist plots of pristine and recrystallized β-Li₃PS₄ (a) and Li₃PS₄·2NMF (b) measured at 25 °C with corresponding fits and Arrhenius plots (c).*

# 4 Conclusions

The study shown here highlights the importance of separating solvent-based synthesis from dissolution and recrystallization strategies of β-Li$_3$PS$_4$. Though many solvents have been reported to be suitable for the synthesis of β-Li$_3$PS$_4$, recrystallization requires the use of polar, weakly protic solvents such as NMF. The recrystallization proceeds via the intermediate phase Li$_3$PS$_4$·2NMF, which needs to be fully removed using an optimized heating process. This step is necessary in order to obtain conductivities of the recrystallized β-Li$_3$PS$_4$ comparable to the pristine material.

We emphasize that the increase in amorphous phase has to be regarded as critical for the use of a recrystallization procedure within a recycling process. Though it leads to comparable room temperature conductivities, it is so far not clear to what extent repeated dissolution and recrystallization might impact the functional electrolyte properties which will be decisive to establish these materials within a circular economy. Furthermore, the role of the electrode materials within the electrode composites needs to be considered which might demonstrate the necessity to adapt the recrystallization to the electrode materials used. This will be reported in a follow-up article.

# 5 Acknowledgements

This work has been funded by German federal state of Hessen (Hessen Agentur, HA-Project Number 848/20-08). LMR. and JJ. acknowledge the financial support by the German Federal Ministry of Research and Education (BMBF), projects 03XP0433D and 03XP0430A (FestBatt). CS and BVL acknowledge financial support by the German Federal Ministry of Research and Education (BMBF), projects 03XP0177B (FestBatt) and the Deutsche Forschungsgemeinschaft via the Cluster of Excellence e-conversion (EXC2089). YI and BG acknowledge the funding from the European Research Council (ERC) under the European Union's Horizon 2020 research and innovation programme (grant agreement No. 865855), the support by the Stuttgart Center for Simulation Science (SimTech), and the support by the state of Baden-Württemberg through bwHPC and the German Research Foundation (DFG) through grant no INST 40/467-1 FUGG (JUSTUS cluster).

# 6 Authors' Contribution

KW, WE and OC conceived and designed the study. KW prepared the samples, measured and analyzed the XRD, SEM, Raman and FTIR spectroscopy data. LR measured and analyzed the XPS data. KW and LMR performed the EIS, and interpreted the data together with AIW under the guidance of JJ. TF measured neutron diffraction. CS measured and analyzed TGA and EGA under the guidance of BVL. YI performed the DFT calculations under the guidance of BG. RED helped with the analysis of diffraction data. KW wrote the manuscript. All authors discussed and revised the work.

# 7 Conflicts of Interest

There are no conflicts of interest to declare.

Supplementary Information for

# Dissolution and Recrystallization Behavior of Li$_3$PS$_4$ in Different Organic Solvents

Kerstin Wissel[a,*], Luise M. Riegger[b,c], Christian Schneider[d], Aamir I. Waidha[a], Theodosios Famprikis[e], Yuji Ikeda[f], Blazej Grabowski[f], Robert E. Dinnebier[d], Bettina V. Lotsch[d,g], Jürgen Janek[b,c], Wolfgang Ensinger[a] and Oliver Clemens[h]

[a] Technical University of Darmstadt, Institute for Materials Science, Materials Analysis, Alarich-Weiss-Straße 2, 64287 Darmstadt, Germany

[b] Justus-Liebig-University Gießen, Institute for Physical Chemistry, Heinrich-Buff-Ring 17, 35392 Gießen, Germany

[c] Justus-Liebig-University Gießen, Center for Materials Research (ZfM), Heinrich-Buff-Ring 17, 35392 Gießen, Germany

[d] Max Planck Institute for Solid State Research, Heisenbergstraße 1, 70569 Stuttgart, Germany

[e] Delft University of Technology, Department of Radiation Science and Technology, Mekelweg 15, Delft 2629JB, The Netherlands

[f] University of Stuttgart, Institute for Materials Science, Materials Design, Pfaffenwaldring 55, 70569 Stuttgart, Germany

[g] Ludwig-Maximilians-Universität München, Department of Chemistry, Butenandtstraße 5-13, 81377 München, Germany

[h] University of Stuttgart, Institute for Materials Science, Chemical Materials Synthesis, Heisenbergstraße 3, 70569 Stuttgart, Germany

Corresponding Author:
Dr. Kerstin Wissel
Email: kerstin.wissel@tu-darmstadt.de
Fax: +49 6151 16-21991

Table S 1: Physical and chemical properties of solvents used in this study. [1] Additionally, the water contents of the solvents after drying as determined from Karl-Fischer titration are given as a mean of three measurements.

| Solvent | Classification | Dielectric contant $\varepsilon_r$ | Polarity index | Boiling point $T_{bp}$ [°C] | Water content after drying [ppm] |
|---|---|---|---|---|---|
| Hexane | Non-polar | 1.88 | 0.009 | 68.7 | Not detectable |
| Toluene | Non-polar | 2.38 | 0.099 | 110.6 | 6.2 |
| Tetrahydrofuran | Aprotic, polar | 7.58 | 0.207 | 66.0 | 79.7 |
| Ethyl acetate | Aprotic, polar | 6.02 | 0.228 | 77.2 | 18.6 |
| 1,2-Dimethoxyethane | Aprotic, polar | 7.20 | 0.231 | 84.6 | 12.7 |
| Acetonitril | Aprotic, polar | 35.94 | 0.460 | 81.6 | 88.8 |
| N-Methylformamide | Protic, polar | 182.40 | 0.722 | 200 | 17.7 |
| Isoproponal | Protic, polar | 19.92 | 0.546 | 82.2 | 58.3 |
| Ethanol | Protic, polar | 24.55 | 0.654 | 78.3 | 147.0 |
| Methanol | Protic, polar | 32.66 | 0.762 | 64.5 | 99.2 |

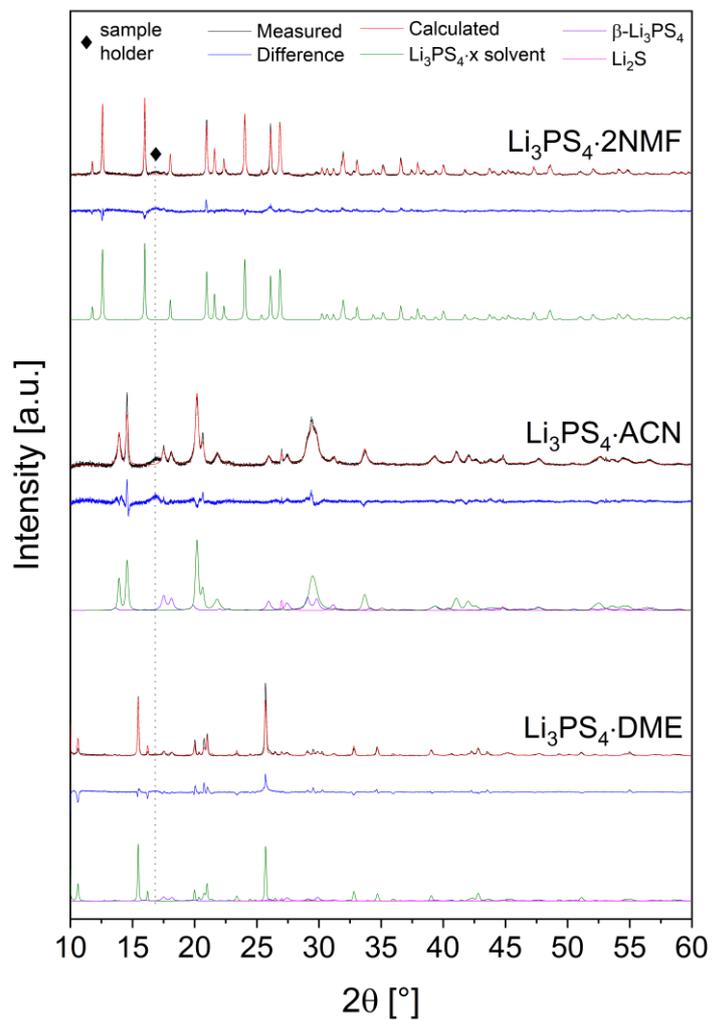

*Figure S1: Rietveld refinements of complexes Li$_3$PS$_4$·DME, Li$_3$PS$_4$·ACN and complexes Li$_3$PS$_4$·2NMF obtained after solvent removal at 120 °C for 4 h under reduced pressure.*

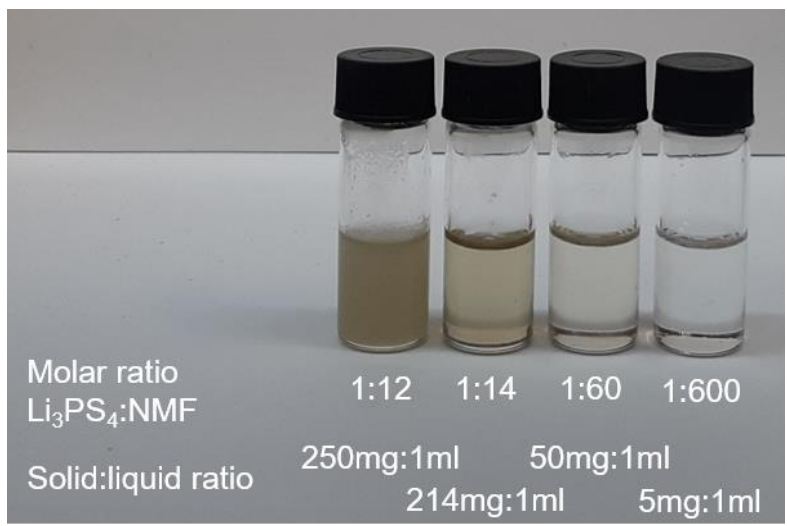

*Figure S2: Photograph of β-Li$_3$PS$_4$ in NMF when using different molar or liquid-to-solid ratios of β-Li$_3$PS$_4$ to NMF.*

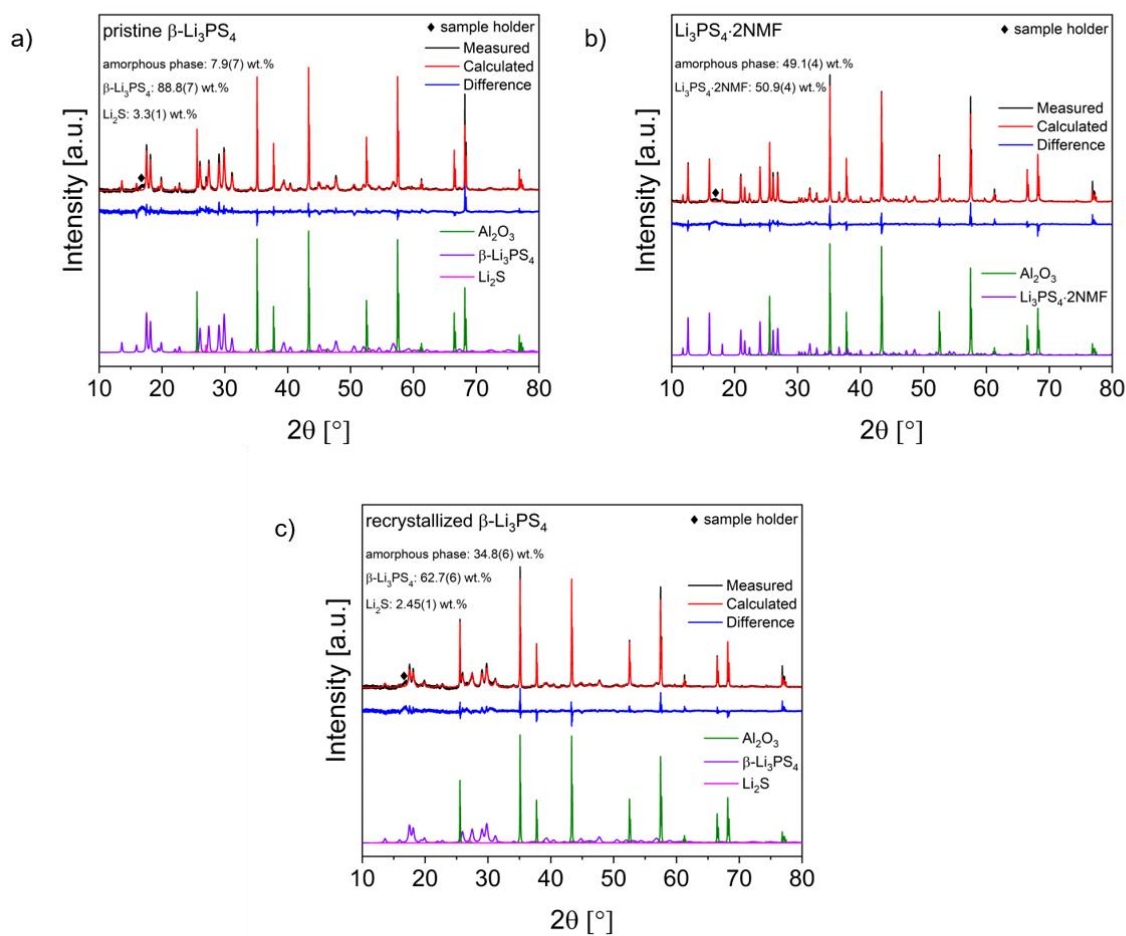

*Figure S3: Rietveld refinements of pristine β-Li$_3$PS$_4$, Li$_3$PS$_4$·2NMF and recrystallized β-Li$_3$PS$_4$ mixed with 50 wt.% Al$_2$O$_3$. From this, using the internal standard method, amorphous phase fractions were determined.*

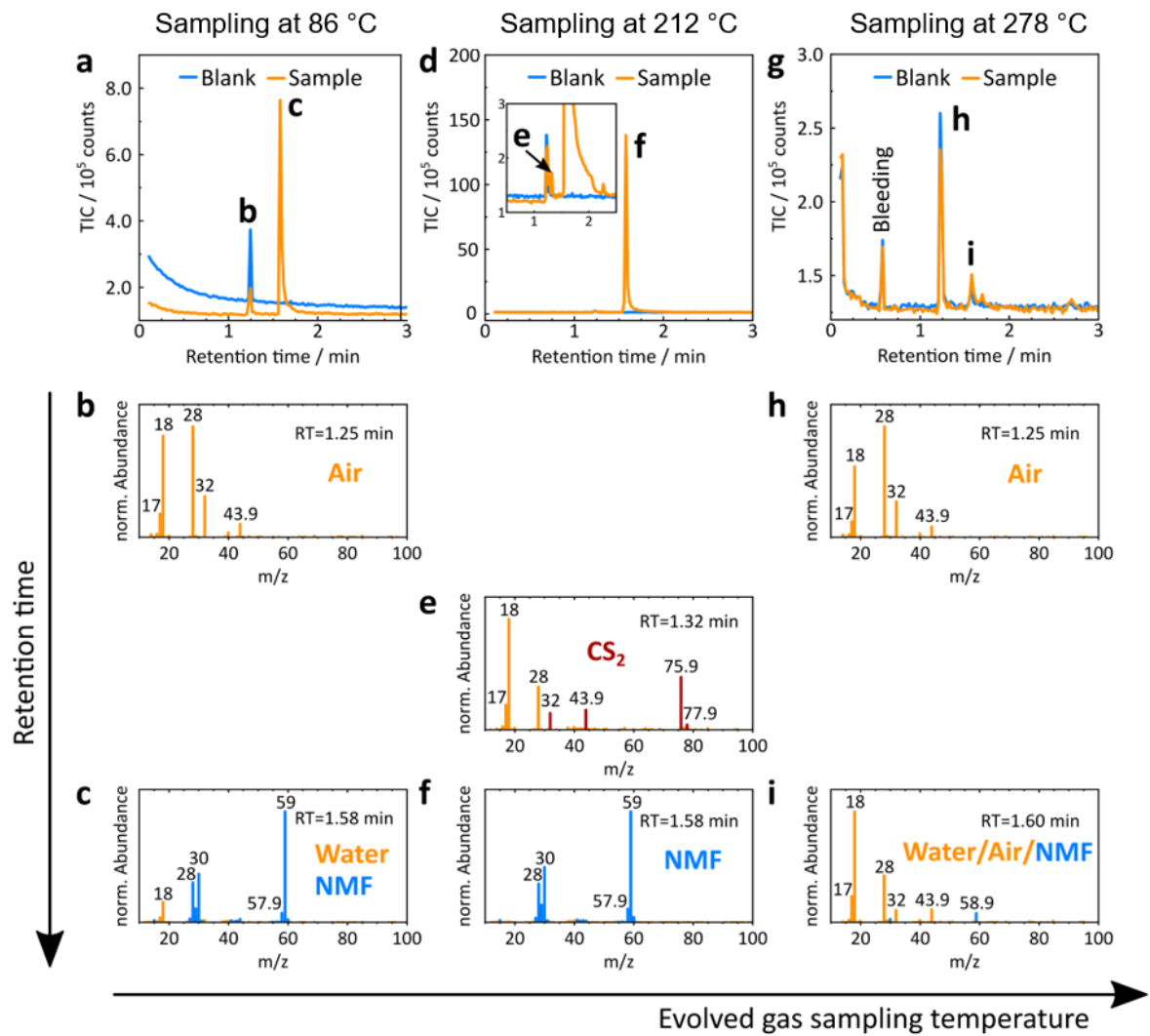

*Figure S4: Evolving gas analysis at sampling temperatures of 86, 212 and 278 °C in dependence of the retention time.*

Table S2: Structural parameters of Li$_3$PS$_4$·2NMF optimized in the DFT simulation.

| Atom | Wyckoff position | x | y | z | Occupancy |
|---|---|---|---|---|---|
| Li1 | 4e | 0 | 0.35097039 | 3/4 | 1 |
| Li2 | 8f | 0.82862866 | 0.18275187 | 0.07521074 | 1 |
| P1 | 4e | 0 | 0.85005548 | 3/4 | 1 |
| S1 | 8f | 0.88733482 | 0.04838892 | 0.71203580 | 1 |
| S2 | 8f | 0.01907986 | 0.65154727 | 0.85591918 | 1 |
| C1 | 8f | 0.60590423 | 0.72475676 | 0.02173654 | 1 |
| C2 | 8f | 0.69529914 | 0.41210385 | 0.11013639 | 1 |
| H1 | 8f | 0.71702934 | 0.31900926 | 0.17289997 | 1 |
| H2 | 8f | 0.61810518 | 0.59983487 | 0.14608463 | 1 |
| H3 | 8f | 0.66208918 | 0.78031623 | 0.01009642 | 1 |
| H4 | 8f | 0.57411478 | 0.86618704 | 0.03581852 | 1 |
| H5 | 8f | 0.55537664 | 0.64208459 | 0.96064680 | 1 |
| N1 | 8f | 0.63972411 | 0.57817329 | 0.09881187 | 1 |
| O1 | 8f | 0.72318715 | 0.36013462 | 0.05485593 | 1 |

**a** [Å] 16.67252033    **b** [Å] 6.08178941    **c** [Å] 16.90241299    **β** [°] 118.09445339

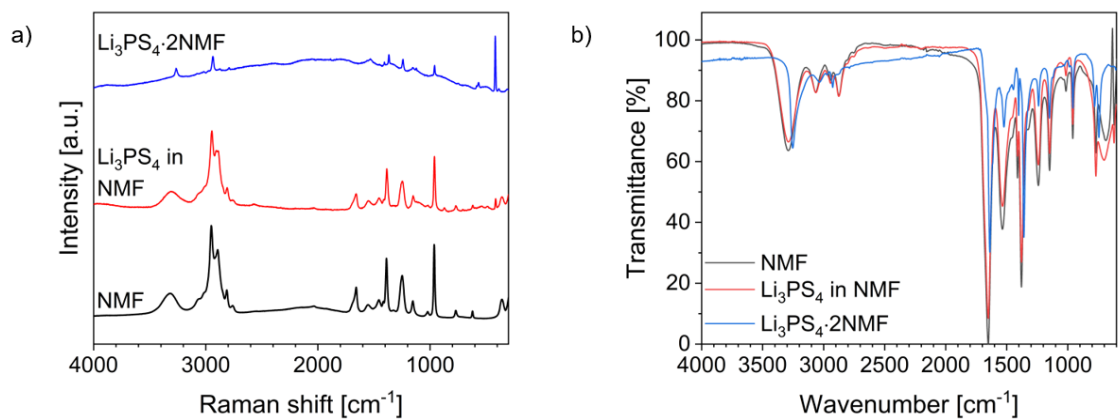

Figure S5: Raman (a) and FTIR (b) spectra of NMF, β-Li$_3$PS$_4$ dissolved in an excess of NMF and Li$_3$PS$_4$·2NMF. The Raman spectrum of β-Li$_3$PS$_4$ dissolved in an excess of NMF is dominated by the spectra of NMF, nevertheless the signal at ~ 421 cm$^{-1}$ corresponding to PS$_4^{3-}$ is still visible, indicating that the ortho-thiophosphate units remain intact in solution. Only after the removal of the excess NMF, the PS$_4^{3-}$ signal becomes dominant. A similar observation can be made for the FTIR spectra where also only after the removal of excess NMF, significant shifts in the bands are observed which correspond well to the observed interactions between Li$_3$PS$_4$ und NMF in Li$_3$PS$_4$·2NMF.

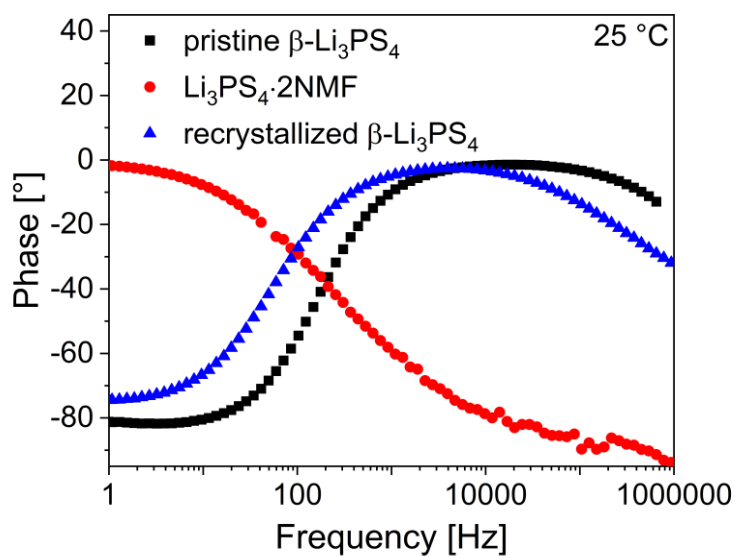

*Figure S6: Bode plots of pristine β-Li$_3$PS$_4$, Li$_3$PS$_4$·2NMF and recrystallized β-Li$_3$PS$_4$.*

# References

1. Appendix A. Properties, Purification, and Use of Organic Solvents. In *Solvents and Solvent Effects in Organic Chemistry*, 2010; pp 549-586.